\documentclass[prb,nofootinbib,superscriptaddress,twocolumn,floatfix]{revtex4-1}
\usepackage{graphicx}
\usepackage{epstopdf}
\usepackage{amsmath}
\usepackage{dcolumn}

\usepackage[normalem]{ulem}

\newcommand{\beq}{\begin{equation}}
\newcommand{\eeq}{\end{equation}}
\newcommand{\bea}{\begin{array}}
\newcommand{\eea}{\end{array}}
\newcommand{\bey}{\begin{eqnarray}}
\newcommand{\eey}{\end{eqnarray}}

\usepackage{color}

\begin{document}

\title{The Coulomb potential in quantum mechanics revisited}
\author{A.A. Othman}
\email{aothman@taibahu.edu.sa. \ \ \ \ Present Address: Department of Physics, University of Waterloo 200 University Avenue West, Waterloo, ON, N2L 3G1, Canada}
\affiliation{Dept. of Physics, University of Alberta, Edmonton, AB, Canada T6G 2E1,\\ {\rm and} \\
Department of Physics, Faculty of Science, Taibah University, Al Madinah Al Munawwarah, Saudi Arabia}
\author{M. de Montigny}
\email{mdemonti@ualberta.ca}
\affiliation{Facult\'e Saint-Jean, University of Alberta, Edmonton, AB, Canada T6C 4G9}
\author{F. Marsiglio}
\email{fm3@ualberta.ca}
\affiliation{Department of Physics, University of Alberta, Edmonton, AB, Canada T6G 2E1}

\begin{abstract}
The procedure commonly used in textbooks for determining the eigenvalues and eigenstates for a particle in an attractive 
Coulomb potential is not symmetric in the way the boundary conditions at $r=0$ and $r \rightarrow \infty$ are considered. 
We highlight this fact by solving a model for the Coulomb potential with a cutoff 
(representing the finite extent of the nucleus); in the limit that the cutoff is reduced to zero we recover the standard result, 
albeit in a non-standard way. This example is used to emphasize that a more consistent approach to solving the Coulomb
problem in quantum mechanics requires an examination of the non-standard solution. The end result is, of course, the same.
\end{abstract}

\date{\today} 
\maketitle

\section{introduction}

The solution of the quantum mechanical problem of determining the energy levels of a (bound) particle in the presence of 
an attractive Coulomb potential, i.e. the hydrogen atom with centre-of-mass coordinate removed, was a spectacular achievement by
Schr\"odinger, published in the same paper in which his famous equation was first introduced,\cite{schrodinger26a} early in 1926.
This solution is now reproduced in every undergraduate textbook on quantum mechanics, with additional steps inserted to make the
derivation easier to understand for the novice. The purpose of this note is to draw attention to the omission of an important part of this
derivation; including it of course ultimately necessarily leads to the same result, with the consequence that the problem is addressed in
what we consider a more systematic manner.

We will first summarize the standard process for the Coulomb potential, mostly in words; the detailed mathematics is available in many textbooks, of which several
clearly laid out ones are cited here.\cite{griffiths05,shankar94,gasiorowicz96,cohen-tannoudji77,bransden00,townsend12} As
described below, all of these references use a power series solution that requires truncation to avoid an un-normalizable solution
{\it{at $r \rightarrow \infty$}}. The other solution is assumed to diverge as $r \rightarrow 0$, and is 
discarded for that reason (but it will be shown that there are certain energy values for which the second solution does not diverge 
as $r \rightarrow 0$.) We will
demonstrate that the symmetric equivalent of this procedure is also possible --- discard the solution that 
diverges as $r \rightarrow \infty$, and truncate the other solution to avoid a divergence as $r \rightarrow 0$. To highlight this second procedure, we consider a more 
realistic problem, the Coulomb potential with a cutoff
near the origin, where we are forced to follow this route to the solution. This problem is anyways more physical than the pure Coulomb problem, as this cutoff models the finite extent of the nucleus. While this necessarily requires a 
knowledge of more complicated mathematical functions, it can be argued that a rudimentary knowledge of this mathematics 
is necessary to fully appreciate even the standard Coulomb problem, where both procedures are possible, and students have
a choice on how to proceed.

\section{The Textbook Coulomb Problem}

The standard treatment is as follows.\cite{griffiths05} The Hamiltonian for the Coulomb potential is given by
\begin{equation}
H = -{\hbar^2 \over 2m}\nabla^2 - {e^2 \over 4 \pi \epsilon_0}{1 \over r},
\label{ham}
\end{equation}
with the first and second terms representing the kinetic and potential energies, respectively of a particle with mass $m$
(this is the reduced mass of the electron if this Hamiltonian arises from the hydrogen problem). Since the Coulomb potential is
central, the solution for the angular part of the wave function is standard, and one is left with the radial equation.
The radial equation for $u(\rho) \equiv r R(r)$, where $R(r)$ is the radial part
of the wave function and $\rho \equiv \kappa r$, with $\kappa \equiv \sqrt{(-2mE)}/\hbar$, is usually rendered in dimensionless form; it is
given by
\begin{equation}
{d^2u \over d\rho^2} = \left[1 - {\rho_0 \over \rho} + {\ell (\ell + 1) \over \rho^2} \right] u.
\label{uofrho}
\end{equation}
Here, $\ell$ is the azimuthal quantum number,
$\rho_0 \equiv 2/(\kappa a_0)$ with $a_0 \equiv 4\pi \epsilon_0 \hbar^2/(me^2)$ the Bohr radius,
and $E<0$ indicates that we are considering bound states. Asymptotic solutions are then `peeled off'
by examining the behavior as $\rho \rightarrow \infty$ and $\rho \rightarrow 0$.
A more general consideration rules out solutions that diverge at the origin; when this is addressed at all (e.g. see Sect.~12.6 in
Ref.~\onlinecite{shankar94}), it is based on normalization and/or conditions of hermiticity. However, the elimination of such solutions
on general grounds is premature in some cases, as will become evident in the next section. 

Incorporating the asymptotic behavior, one writes the solution $u(\rho)$ as
\begin{equation}
u(\rho) = \rho^{\ell + 1} e^{-\rho} v(\rho),
\label{ansatz}
\end{equation}
and writes a new 2nd order differential equation for $v(\rho)$ (see below). This is then solved in one of two ways: (i) most commonly this
function is expanded in a power series in $\rho$ and then a recursion relation is derived for the coefficients in the power series,
or (ii) the equation is recognized as the differential equation for the confluent hypergeometric function, and then the solution is
simply written down as the Kummer function.\cite{landau77,nist10} 
\begin{figure}[h]
\center
\includegraphics[scale=0.40]{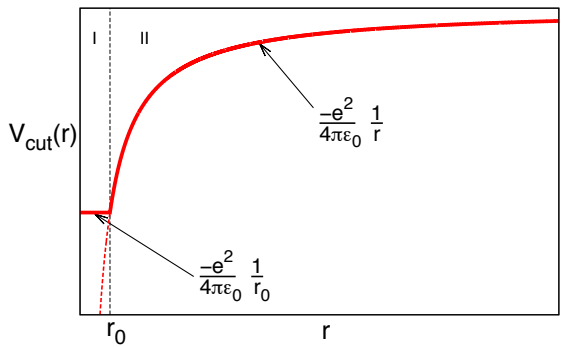}
\caption{$V_{\rm cut}(r)$ vs. $r$, as given in Eq.~(\ref{VCcut}). The range of the `nuclear' part, $r_0$ is highly exaggerated
in the figure.}
\label{hfig1}
\end{figure}
In either case it is recognized that in fact the Kummer function
diverges as $e^{2\rho}$ as $\rho \rightarrow \infty$, which overwhelms the `peeled-off' solution, and gives rise to a non-normalizable
wave function. In the version that utilizes the power series, a remedy is then recognized: by making one of the parameters in the
problem, $\rho_0 \equiv 2/(\kappa a_0)$, equal to a positive even integer, the recursion relation is truncated, so instead of an
infinite power series that describes exponentially growing behavior, we obtain a polynomial of finite order.
The same conclusion is reached for those familiar with the properties of the Kummer function, and in fact one recognizes that
these polynomials are the Associated Laguerre polynomials.\cite{nist10,abramowitz72}
The radial part of the wave function therefore consists of an Associated Laguerre polynomial times an exponential 
with argument $-r/(a_0n)$ and
$n$ is a positive integer, the so-called principal quantum number.

\section{Bound-state solutions for the Coulomb potential with a cutoff near the origin\label{Coulomb}}

Because the radius of a proton is of order one femtometer, roughly five orders of magnitude smaller than the Bohr radius, it is
usually disregarded (except perhaps as an example of a perturbation) in undergraduate studies of the hydrogen atom.
Nonetheless, a more realistic potential for the hydrogen atom is
\beq
V_{\rm cut}(r)=\left\{ \begin{array}{lll}
-\frac {e^2}{4 \pi \epsilon_0}{1 \over r_0}, \qquad & 0<r\leq r_0, & ({\rm region}\ I)\\ 
 & & \\
-\frac {e^2}{4 \pi \epsilon_0}{1 \over r},\qquad & r\geq r_0, & ({\rm region}\ II)\end{array}\right.
\label{VCcut}\eeq
where $r_0$ represents the radius of the nucleus. A schematic is provided in Fig.~\ref{hfig1}. 
One immediate question a novice might ask is, does this potential support an infinite
number of bound states as is the case for the Coulomb potential without a cutoff? As we shall see below, the answer is `yes,'
obvious to those who realize this infinite number of bound states is associated with the long-range
tail of the Coulomb potential (and not with the singular behavior near the origin).
The strategy for the solution to this problem is standard; determine solutions appropriate to the two regions, with
arbitrary coefficients, and then match the wave function and its derivative at $r=r_0$ to determine the remaining
coefficients.

With $\ell = 0$ the solution for $0< r< r_0$ is elementary --- a linear combination of $\sin{(qr)}$ and $\cos{(qr)}$ with the coefficient of
the $\cos{(qr)}$ solution set to zero to achieve the proper behavior at $r=0$ (i.e. $u(r) = 0$ as $r \rightarrow 0$),
with $q \equiv \sqrt{{2m}(E + V_0)/\hbar^2}$,
and $V_0 \equiv e^2/(4 \pi \epsilon_0 r_0)$. Therefore, in region I, 
\begin{equation}
u_I(r) = A{\sin{(qr)} \over \sin{(qr_0)}},
\label{region1}
\end{equation}
where $A$ is an unknown coefficient. The solution for $r_0 < r < \infty$ is more difficult. One can attempt a power series in $\rho$,
as was done in the case with no cutoff, and in fact this is the first hint that perhaps the recipe provided in the previous section is
not the whole story. For one thing, it has likely occurred to the reader already that the standard power series solution represents one
solution; since the equation is a 2nd order differential equation, there should be two independent solutions. 
In fact, the equation for $v(\rho)$ follows from insertion of Eq.~(\ref{ansatz}) into Eq.~(\ref{uofrho}) 
\begin{equation}
\rho {d^2v \over d\rho^2} + 2 (\ell + 1 - \rho){dv \over d\rho} +   [\rho_0 - 2(\ell + 1)]v = 0,
\label{hyper}
\end{equation}
and is a particular example of the confluent hypergeometric equation: 
\begin{equation}
z\frac{d^2y}{dz^2}+\left(b-z\right)\frac{dy}{dz}-ay=0,
\label{Kummer}
\end{equation}
whose general solution is
\begin{equation}
y=C\ M\left(a,b,z\right)+D\ U\left(a,b,z\right),
\label{KummerSol}
\end{equation}
where $C$ and $D$ are arbitrary constants. $M\left(a,b,z\right)$ is known as the Kummer confluent hypergeometric 
function, and $U\left(a,b,z\right)$ is known as the Tricomi confluent hypergeometric function; these two solutions are independent
of one another. They are further discussed in the Appendix.
Henceforth we will focus on $\ell = 0$ to simplify the analysis. If we substitute $z\equiv 2\rho$ into Eq.~(\ref{hyper}) then we see from Eq. (\ref{KummerSol}) that this equation has two independent solutions,
\begin{equation}
v(\rho) = C \ M(1 - \rho_0/2,2,2\rho) + D \ U(1 - \rho_0/2,2,2\rho),
\label{two_solutions}
\end{equation}
with $a \equiv 1 - \rho_0/2$ and $b \equiv 2$. Usually, in the confluent hypergeometric functions, $a$ and $b$ are thought of as parameters and $z$ is the variable. It turns out (students are not told this!) the Tricomi function generally diverges as $z \rightarrow 0$ (more on this later).
Perhaps for this reason it is usually not considered in the solution to the usual Coulomb problem.

\begin{figure}[h]
\center
\includegraphics[scale=0.50,angle=-90]{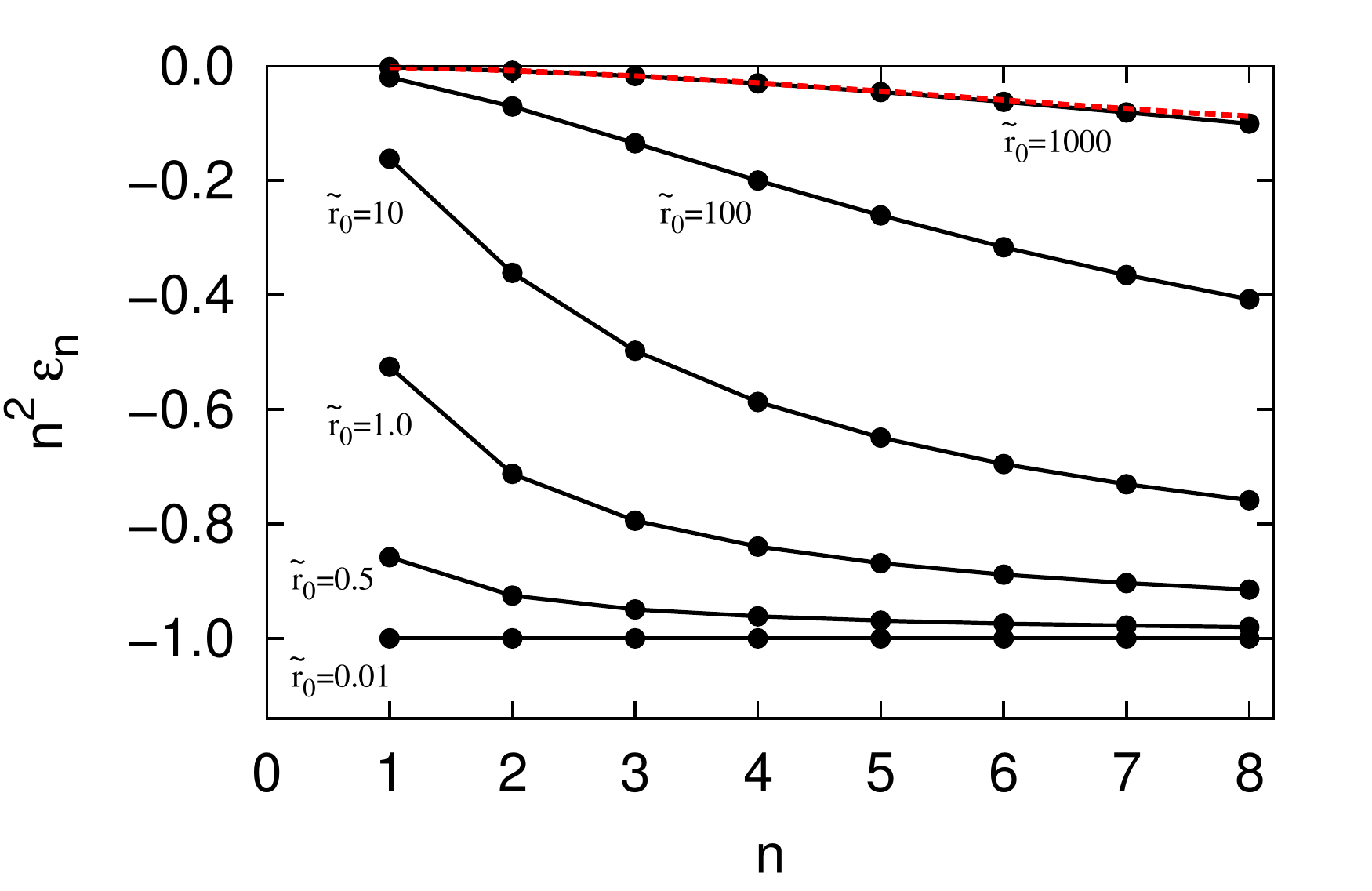}
\caption{The product $n^2 \frac{E_n}{E_0}$ as a function of $n$ for various values of $\tilde{r}_0$, as shown. 
For $\tilde{r}_0$ approaching zero we obtain a horizontal line at $-1$, which corresponds to the eigenvalues of the 
Coulomb potential. With $\tilde{r}_0 = 0.01$ this limit has clearly been achieved.
For $\tilde{r}_0 >>1$, we expect the first few energies to have values almost equal to those of a spherical potential, as given by Eq.~(\ref{sqwell}), indicated with a dashed (red) curve (after multiplication by $n^2$) for $\tilde{r}_0=1000$. The reasonable
agreement with the data for $\tilde{r}_0=1000$ indicates that this limit has been achieved for the lowest energy levels for
this value of $\tilde{r}_0$.}
\label{hfig3}
\end{figure}

But there is a twist! Note that when we wrote down the solution for region I, we eliminated one of the arbitrary constants by 
examining the boundary
condition at $r=0$ (recall $\rho \equiv \kappa r$). Similarly we now eliminate one of the constants for the solution in region II, by examining
the boundary condition at $\rho \rightarrow \infty$, which immediately gives $C = 0$ (since, as we learned in the standard Coulomb
problem, the Kummer function blows up exponentially in this limit (more on this below), and we cannot `salvage' the solution by making $\rho_0$ equal to a
positive even integer --- instead, it will be determined by the matching at $r=r_0$). We now have the remaining task of matching the wave function and its derivative at $r = r_0$. Using Eq.~(\ref{two_solutions}) (with $C=0$) in Eq.~(\ref{ansatz}) and matching with Eq.~(\ref{region1}),
we obtain two equations,
\begin{equation}
A = D \kappa r_0 e^{-\kappa r_0} U(1-\rho_0/2,2,2\kappa r_0),
\label{eqn1}
\end{equation}
and
\begin{eqnarray}
Aq {\rm cot}(qr_0) &&= u_{II}(r_0)\left[{1 \over r_0} - \kappa\right] \nonumber \\ 
+&& 2D \kappa^2 r_0 e^{-\kappa r_0}  {dU(1-\rho_0/2,2,z) \over dz}|_{z=2\kappa r_0}.
\label{eqn2}
\end{eqnarray}
Dividing the latter equation by the former, and inserting the identity,\cite{nist10}
\begin{equation}
{dU(a,b,z) \over dz} = -aU(a+1,b+1,z),
\label{identity}
\end{equation}
gives us an equation to determine the allowed bound state energies,
\begin{equation}
qr_0 {\rm cot}(qr_0) -1 = -\kappa r_0\left[1 + 2  (1 - \rho_0/2){U(2-\rho_0/2,3,2\kappa r_0) \over U(1-\rho_0/2,2,2\kappa r_0)}
\right].
\label{eigenvalue}
\end{equation}
Equation~(\ref{eigenvalue}) can be rewritten in terms of the dimensionless variables, $\tilde{r}_0 \equiv r_0/a_0$ and
\beq
x \equiv 1/\sqrt{-\epsilon}; \ \ \ \ \ \ \ \epsilon = E/E_0; \ \ \ \ \ \ \ \ \  E_0 \equiv {\hbar^2 \over 2 m a_0^2}.
\label{x_epsilon}
\eeq
The equation becomes
\begin{eqnarray}
&&\sqrt{\tilde{r}_0\left(2-{\tilde{r}_0 \over x^2}\right)}\cot \sqrt{\tilde{r}_0\left(2-{\tilde{r}_0 \over x^2}\right)} -1 = \nonumber \\
&&-{\tilde{r}_0 \over x} \left[1+2\left(1-x\right)\frac{U\left(2-x,3,{2\tilde{r}_0 \over x}\right)}{U\left(1-x,2,{2\tilde{r}_0\over x}\right)}\right].
\label{eqn_in_x}
\end{eqnarray}
Note that we require solutions $x$ as a function of $\tilde{r}_0$ in order to determine the energy. The virtue of using the variable $x$
is that the solutions for $x$ should approach the positive integers as the cutoff $\tilde{r}_0$ approaches zero. 
\begin{figure}[h]
\center
\includegraphics[scale=0.50,angle=-90]{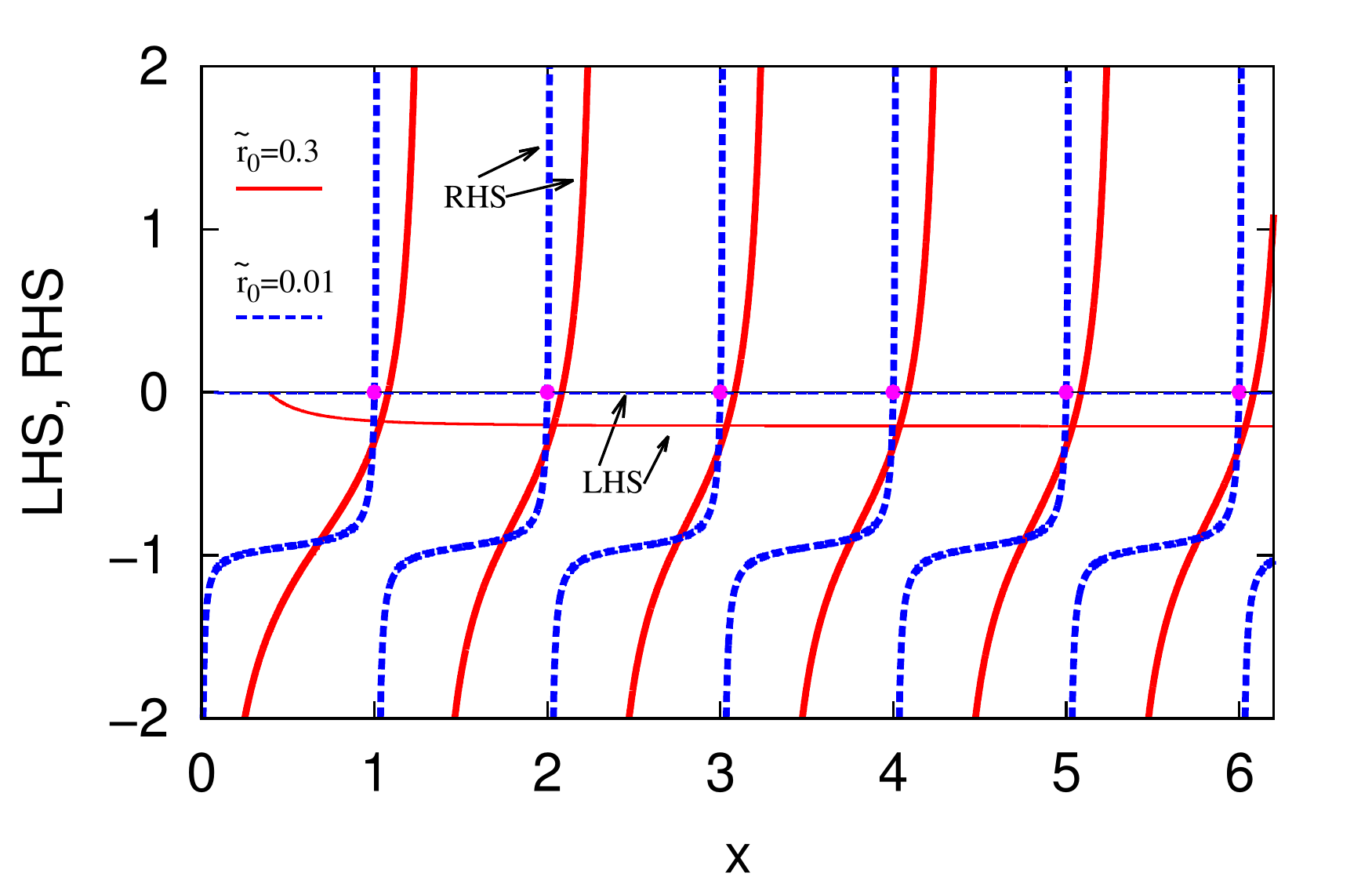}
\caption{The left-hand side (LHS) and right-hand side (RHS) of Eq. (\ref{eqn_in_x}) plotted as a function of $x \equiv 1/\sqrt{-\epsilon}$.
Solid (red) curves are for the parameter $\tilde{r}_0 = 0.3$, while, for reference, we have also plotted the solutions for
$\tilde{r}_0 = 0.01$ [dashed (blue) curves]. In both cases the thicker curves with many branches (two of which are labelled)
refer to the RHS, while the thinner curves (both labeled) refer to the LHS. The solutions to Eq. (\ref{eqn_in_x}) are given by
the intersection of thin and thick curves. For $\tilde{r}_0 = 0.01$ these essentially coincide with the integers, as indicated by the
pink dots, since there is essentially no cutoff. For $\tilde{r}_0 = 0.3$ (red curves) the solutions are clearly at higher values of $x$; given
that $x \equiv 1/\sqrt{-\epsilon}$ this corresponds to higher values of energy, as we would expect.}
\label{hfig2}
\end{figure}
Fig.~\ref{hfig2} illustrates the graphical solution as represented by the left-hand-side (LHS) and right-hand-side (RHS) of
Eq.~(\ref{eqn_in_x}) for two different values of $\tilde{r}_0$. The solutions shown here make apparent that the energies increase
as $\tilde{r}_0$ increases from zero. In Fig.~\ref{hfig3}  we show the solutions for a variety of values of $\tilde{r}_0$ showing 
how the limit of the
Coulomb potential (no cutoff) is approached for sufficiently small values of $\tilde{r}_0$. It is also evident that the number of bound
states remains fixed, i.e. there is a one-to-one correspondence between bound state energies for the Coulomb potential and those
for the Coulomb potential with a cutoff, even if the cutoff is $1000 \times$ the Bohr radius. 

More specifically, as $r_0/a_0$ increases from zero, the energy eigenvalues are all slightly increased in value (reduced in magnitude), 
$\epsilon_n \approx -(1-\delta_n)^2/n^2$, where $\delta_n$ is a small positive quantity. Increasing $r_0$ to values $r_0 >> a_0$
increases these eigenvalues further, but {\it all these states remain bound}. For very large $r_0$ the potential resembles a finite
square well, with (shallow) depth $V_0 \equiv {e^2 \over 4 \pi \epsilon_0}{1 \over r_0}$ and (large) width $r_0$, augmented with a
Coulomb tail. Viewed as an attractive square well potential, the lowest energy levels are given by
\beq
\epsilon_n \approx -{2 \over \tilde{r}_0} + \left( {n \pi \over \tilde{r}_0} \right)^2,
\label{sqwell}
\eeq
so that even in this limit the argument of the cotangent function on the left-hand-side of Eq.~(\ref{eqn_in_x}) remains real. The bound 
state energies as a function of the principal quantum number $n$ are plotted in Fig.~\ref{hfig3} for various values of $\tilde{r}_0$,
where the two limits are clearly indicated. For a
Coulomb potential with no cutoff ($\tilde{r}_0 \rightarrow 0$) we expect all results at $n^2 \epsilon_n = -1$, while the opposite extreme
($\tilde{r}_0 \rightarrow \infty$), the dashed curve is Eq.~(\ref{sqwell}) for $\tilde{r}_0 = 1000$ and indicates that the results have
approached the limit described by Eq.~(\ref{sqwell}).

\section{So What?}

\begin{figure}[h]
\center
\includegraphics[scale=0.5,angle=-90]{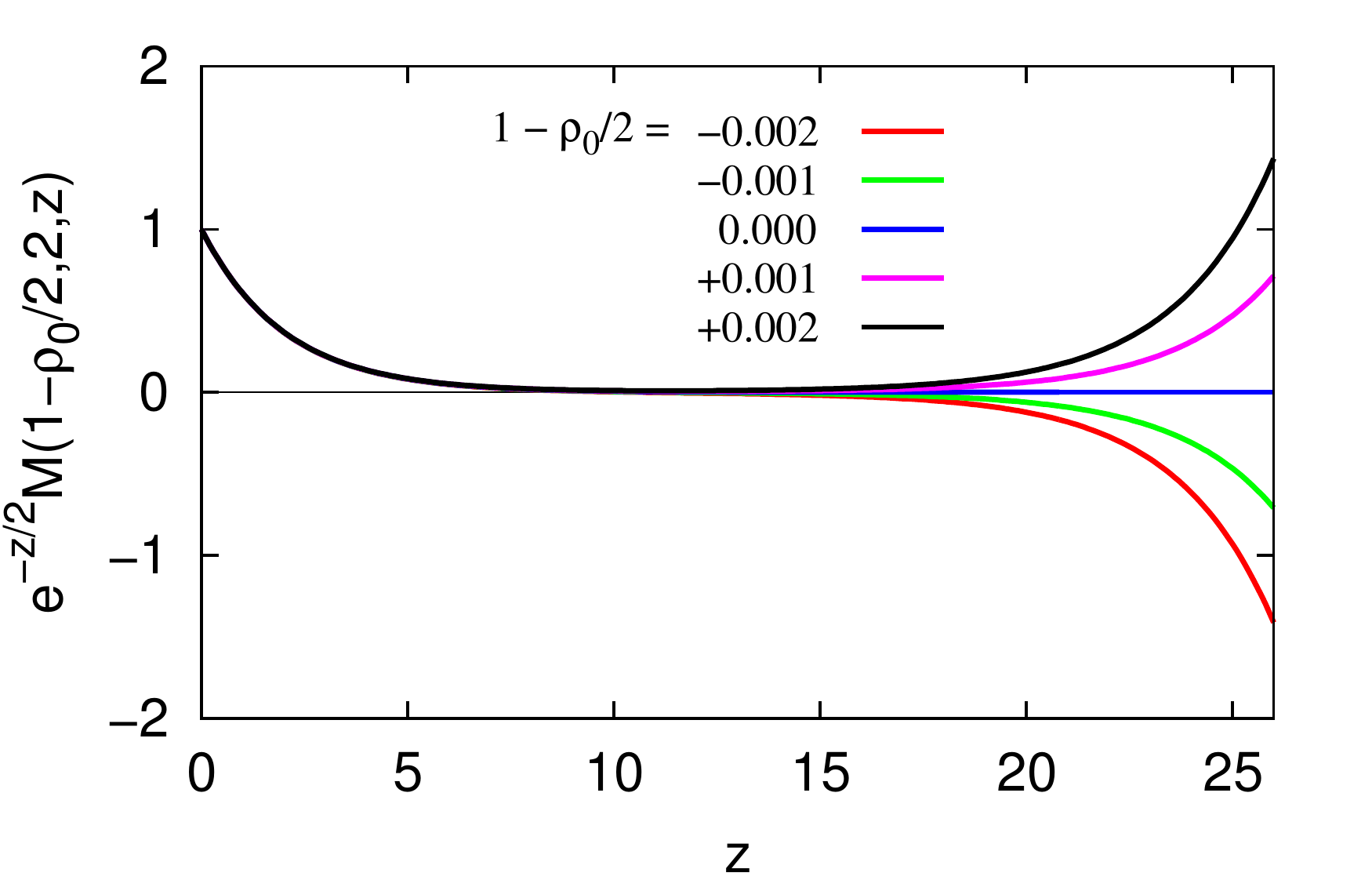}
\caption{The function $e^{-z/2}M(1-\rho_0/2, 2, z)$ as a function of $z$, with $\rho_0=2.004$ (red, lowest),
$\rho_0=2.002$ (green, second lowest), $\rho_0=2.000$ (blue, middle curve), $\rho_0=1.998$ (pink, second highest), 
and $\rho_0=1.996$ (black, highest), where the rankings refer to the far right of the figure. Note that on the left of the
figure the results all agree with one another to high accuracy. Also note that if one investigated this function only out
to $z=10$ or $12$, it would appear to converge for all values of the parameter $1-\rho_0/2$. Only for larger values of
$z$ is it clear that for non-integer values of $1-\rho_0/2$ this function actually diverges. Here, $M(a,b,z)$ is the so-called
Kummer function; details on how to calculate it are given in the Appendix.}
\label{hfig4}
\end{figure}

We have solved for a more realistic variation of the Coulomb potential. If we let $r_0 \rightarrow 0$ we should recover 
the usual results. However, returning to the discussion in the previous section following Eq.~(\ref{two_solutions}) we note
that in this limiting process, we are left only with the Tricomi function, $U(1-\rho_0/2,2,2\rho)$. We know that the Kummer function
will reduce to the Laguerre polynomials (note\cite{remark1} that we use the physicist's definition of the Laguerre
polynomials, as found for example in Ref.~\onlinecite{griffiths05}) through
\begin{equation}
\lim_{\rho_0 \rightarrow 2n} M(1-\rho_0/2,2,2\rho) = {1 \over n}{1 \over n!} L_{n-1}^1(2\rho);
\label{laguerrem}
\end{equation}
but the Kummer function has been eliminated by setting the coefficient $C=0$.
For reference, Fig.~\ref{hfig4} shows the product of $e^{-z/2}$ and the
Kummer function $M(1-\rho_0/2,2,z)$ as a function of $z$ for several values of $\rho_0$ close to $2.0$. This combination
diverges except for the special case when $\rho_0 = 2n$, with $n$ a positive integer, in this case, $n=1$. This is the condition that
normally ``saves'' the solution to the standard Coulomb potential from blowing up and gives us the Coulomb eigenvalues. 
However, in the way we have set up the
problem, this solution is no longer salvageable as $r_0 \rightarrow 0$, as it was eliminated from the start. How are we to recover 
the known solutions?

\begin{figure}[h]
\center
\includegraphics[scale=0.5,angle=-90]{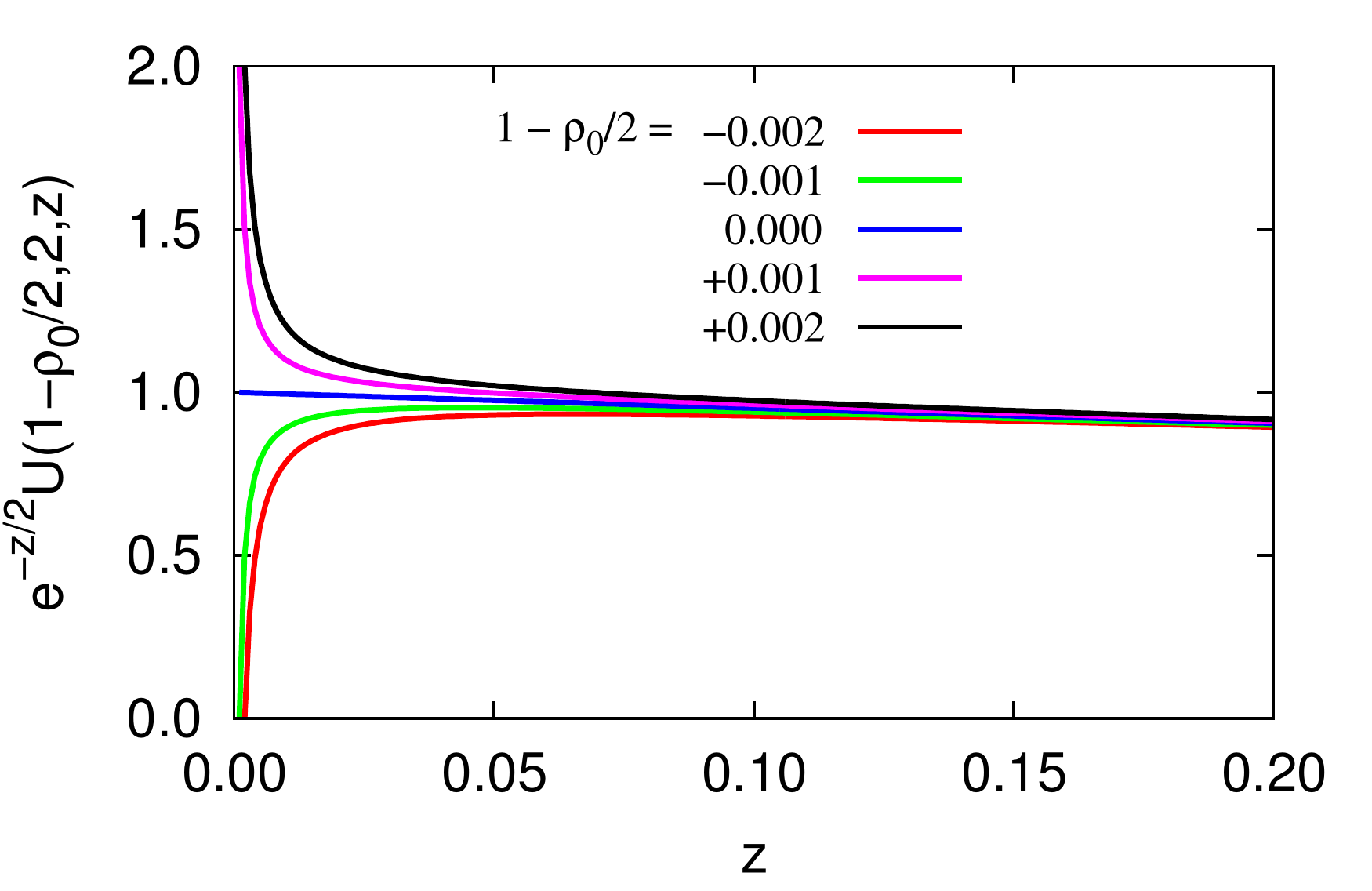}
\caption{The function $e^{-z/2}U(1-\rho_0/2, 2, z)$ as a function of $z$, with $\rho_0=2.004$ (red, lowest),
$\rho_0=2.002$ (green, second lowest), $\rho_0=2.000$ (blue, middle curve), $\rho_0=1.998$ (pink, second highest), 
and $\rho_0=1.996$ (black, highest), where the rankings refer to the far left of the figure. Note that on the right of the
figure the results all agree with one another to high accuracy. It is clear that for non-integer values of $1-\rho_0/2$ this 
function actually diverges as $z \rightarrow 0$. Here, $U(a,b,z)$ is the so-called
Tricomi function, further discussed in the Appendix.}
\label{hfig5}
\end{figure}

The answer is provided in Fig.~\ref{hfig5}, where the product of $e^{-z/2}$ and the Tricomi function $U(1-\rho_0/2,2,z)$ is plotted as
a function of $z$ for several values of $\rho_0$ close to $2.0$. This function is always well behaved as $z \rightarrow \infty$, but 
tends to diverge as $z \rightarrow 0$, {\it except} in the case where $\rho_0 = 2n$, with $n$ an positive integer --- precisely the condition
that yields the known eigenvalues for the Coulomb potential. The case shown in
Fig.~\ref{hfig5} corresponds to $n=1$. It is also true (and probably less known) that
\begin{equation}
\lim_{\rho_0 \rightarrow 2n} U(1-\rho_0/2,2,2\rho) = {(-1)^{n-1} \over n} L_{n-1}^1(2\rho)
\label{laguerreu}
\end{equation}
so we indeed recover not only the correct eigenvalues but also the correct eigenfunctions, when $r_0 \rightarrow 0$.

The point we wish to make is that, even when we consider the usual Coulomb potential, without a cutoff, we should include
the Tricomi solution as well as the Kummer solution. Both are ``saved'' (i.e. rendered normalizable) in the same way, by having
$\rho_0 = 2n$ with $n$ a positive integer. That is, both functions, $M(a,b,z)$ and $U(a,b,z)$, reduce to Laguerre polynomials
when the parameter $a$ is a negative integer (and $b$ is a non-negative integer). There is therefore an
equivalent symmetric procedure for solving this standard problem; one can first view the boundary condition at $r \rightarrow \infty$,
realize that the Kummer function diverges there, and therefore set the constant in front of this function equal to zero, as is
normally done (usually implicitly) for the Tricomi solution. Having done this, one can now declare the Tricomi function to be the
solution, only to discover on more careful examination that this function diverges (and is un-normalizable) as $r \rightarrow 0$. 
We can then discover that this difficulty is overcome by requiring $\rho_0 = 2n$ with $n$ a positive integer, which gives both
the correct eigenvalues and the correct eigenfunctions.

\section{Summary}

We have presented solutions for the cutoff Coulomb potential, a model for the hydrogen atom that includes the finite
extent of the nucleus. The number of bound states remains infinite, on a one-to-one mapping with the solutions for the
standard Coulomb problem. Naturally, they are elevated in value compared to the standard Coulomb problem. To solve
this problem we have followed the procedure normally followed for the standard problem, except it has been necessary
to include the two independent solutions to the radial equation. We have further shown that this more difficult procedure
can also be followed for the standard problem. That is, either the Kummer function
{\it or} the Tricomi function can be
retained in the solution to the standard problem. Both these functions cause difficulties; the former diverges at $r \rightarrow \infty$, while
the latter diverges at $r=0$. Divergences at both ends, near $r=0$ and for $r\rightarrow \infty$ are
prevented by a quantization condition which is identical at either end, and ultimately gives the usual Coulomb eigenvalues,
$E = -E_0/n^2$, with $E_0 = \hbar^2/(2ma_0^2)$, with the usual eigenstates, proportional to the Laguerre polynomials. The usual
procedure only recognizes the `salvaging' of the Kummer solution; one of the primary purposes of this paper is to alert instructors
and students that for the Coulomb potential both solutions are possible and an equivalent symmetric
procedure is available, as outlined here. The standard
procedure for `salvaging' the one (demanding that $\rho_0 = 2n$ where $n$ is a positive integer) also `salvages' the other. Therefore
the correct eigenvalues and eigenvectors are obtained in either case.

\section*{Acknowledgements}
A. Othman acknowledges financial support from the Taibah University (Medina, Saudi Arabia).
We are also grateful to the Natural Sciences and Engineering Research Council of Canada (NSERC), to the Alberta iCiNano program, and to the University of Alberta Teaching and Learning Enhancement Fund (TLEF) grant
for partial support.

\section*{Appendix}

Two independent solutions to the confluent hypergeometric equation (also sometimes called Kummer's Equation),
\begin{equation}
z\frac{d^2y}{dz^2}+\left(b-z\right)\frac{dy}{dz}-ay=0,
\label{Kummer_app1}
\end{equation}
are given by the Kummer function, $M(a,b,z)$, and the Tricomi function, $U(a,b,z)$.\cite{nist10,abramowitz72}
While these are not familiar to 
most undergraduates, they underly the known (and correct) solution to the bound and excited eigenstates of the 
single particle problem in a Coulomb potential. They each have a number of representations; for the Kummer function,
a power series solution is given by
\begin{equation}
M(a,b,z) = 1 + {a \over b} z + {a(a+1) \over b(b+1)} {z^2 \over 2 !} + {a(a+1)(a+2) \over b(b+1)(b+2)}{z^3 \over 3!} + ....,
\label{kummer_app2}
\end{equation}
which exists for all parameter and variable values except when $b$ is a non-positive integer. Eq.~(\ref{kummer_app2}) is
written more concisely, using so-called Pochammer symbols, $(a)_k$, where
\begin{equation}
(a)_k \equiv a(a+1)(a+2)....(a+k-1), \phantom{aaaaaaaa} (a)_0 \equiv 1.
\label{poch}
\end{equation}
These are simple if $a$ is an integer. For example, $(2)_k = (k+1)!$, $(3)_k = (k+2)!/2$, and so on. The concise form is then
\begin{equation}
M(a,b,z) = \sum_{k=0}^\infty {(a)_k \over (b)_k} {z^k \over k!}.
\label{kummer_sum}
\end{equation}
Note that this function
has simple limiting forms,
\begin{equation}
\lim_{z \rightarrow 0} M(a,b,z) = 1,
\label{kummer_limits}
\end{equation}
and
\begin{equation}
\lim_{z \rightarrow \infty} M(a,b,z) = e^z z^{a-b}/\Gamma(a), \ \ \ a \ne 0, -1, -2, ....
\label{kummer_limits2}
\end{equation}
where $\Gamma(a)$ is the Gamma Function.\cite{abramowitz72,nist10} Notice that $M(a,b,z)$ is generally divergent as $z$
increases, except (refer back to Eq.~(\ref{kummer_app2})) if $a$ is equal to a negative integer. Then in fact the infinite series
terminates, and $M(a,b,z)$ becomes a polynomial. In fact this is already quoted in the text, and we repeat Eq.~(\ref{laguerrem}) here
in more generic form, for the case encountered in the Coulomb problem ($b=2$):
\begin{equation}
M(-n,2,z) = {1 \over n+1}{1 \over (n+1)!} L_{n}^1(z), \ \  n=0, 1, 2, ....
\label{laguerre_gen}
\end{equation}
and the polynomial is identified as the Associated Laguerre polynomial.\cite{remark1} Note that the terms in the summation
Eq.~(\ref{kummer_sum}) satisfy a recursion relation,
\begin{equation}
M(a,b,z) = \sum_{k=0}^\infty S_k, \ \ {\rm with}\ \ S_{k} = {(a+k-1) z  \over k(b+k-1)}S_{k-1},
\label{recursion}
\end{equation}
which makes Eq.~(\ref{kummer_sum}) very easy to program. Convergence is very fast; for everything required in this
manuscript, 30 terms in the summation were more than enough for 8-digit accuracy.

Much of this will be somewhat familiar
to the student who has studied the power series solution for the Coulomb potential --- it is just Eq.~(\ref{kummer_app2}), and requiring
the parameter $a$ to be a non-positive integer is precisely the condition required to `salvage' this solution, i.e. to keep it normalizable.

The Tricomi function is less familiar. The power series solution is, when $b=n+1$, $n=0, 1, 2, ....$, and $a \ne 0, -1, -2, ...$,
\begin{eqnarray}
U(a,n+1,z) &=& {(-1)^{n+1} \over n!\Gamma(a-n)} \sum_{k=0}^\infty {(a)_k \over (n+1)_k} {z^k \over k!}
h(k,n,a,z) \nonumber \\
&+& {1 \over \Gamma(a)} \sum_{k=1}^n {(k-1)!(1-a+k)_{n-k} \over (n-k)!}{1 \over z^k}
\label{tricomi_sum}
\end{eqnarray}
where $\psi(z)$ is the Digamma function\cite{abramowitz72} and 
\begin{equation}
h(k,n,a,z) \equiv  {\rm ln} z + \psi(a+k) - \psi(k+1) - \psi(n+k+1) \nonumber.
\label{defn_of_h}
\end{equation}
For $a=-m$, $m = 0, 1, 2, ...$,
\begin{equation}
U(-m,n+1,z) =(-1)^{m} \sum_{k=0}^m \left( {m \atop k}\right) (n+k+1)_{m-k}(-z)^k,
\label{tricomi_sum2}
\end{equation}
with $\left( {m \atop k}\right) \equiv {m! \over k! (m-k)!}$.
These are forms we have found suitable for programming, again with no more than 30 terms required for high accuracy. Similar to
Eq.~(\ref{laguerre_gen}), a special case of Eq.~(\ref{tricomi_sum2}) pertinent to the Coulomb potential is
\begin{equation}
U(-n,2,z) = {(-1)^n \over n+1} L_{n}^1(z), \ \  n=0, 1, 2, ....
\label{laguerre_gen2}
\end{equation}
Once again, the Associated Laguerre polynomials appear, this time as a result of `salvaging' the solutions that would
otherwise diverge at the origin. Thus, for special parameter values ($b=2$ and $a=-n$, $n = 0, 1, 2, ...$) both (formerly)
independent solutions become proportional to the same Associated Laguerre polynomial.

\end{document}